\documentclass[prl,showpacs,showkeys,preprintnumbers,amsmath,amssymb,superscriptaddress,twocolumn]{revtex4}
\usepackage{graphicx}
\usepackage{amssymb,amsfonts,amsmath}
\usepackage{epstopdf}
\usepackage{color}
\usepackage{subfigure}
\usepackage{float}

\def\lmax{\ell_{\rm max}}

\def\N{{\mathcal N}}

\begin{document}

\title{How anisotropy beats fractality in two-dimensional on-lattice DLA growth}

\author{Denis~S.~Grebenkov}
 \email{denis.grebenkov@polytechnique.edu}

\affiliation{Laboratoire de Physique de la Mati\`{e}re Condens\'{e}e (UMR 7643), \\ 
CNRS -- Ecole Polytechnique, University Paris-Saclay, 91128 Palaiseau, France}

\affiliation{Interdisciplinary Scientific Center Poncelet (ISCP),%
\footnote{International Joint Research Unit -- UMI 2615 CNRS/ IUM/ IITP RAS/ Steklov MI RAS/ Skoltech/ HSE, Moscow, Russian Federation} \\
Bolshoy Vlasyevskiy Pereulok 11, 119002 Moscow, Russia}

\author{Dmitry~Beliaev}
 \email{belyaev@maths.ox.ac.uk}

\affiliation{Mathematical Institute, University of Oxford, \\
Andrew Wiles Building, Radcliffe Observatory Quarter, Woodstock Road, Oxford, OX2 6GG, UK}

\date{Received: \today / Revised version: }

\begin{abstract}
We study the fractal structure of Diffusion-Limited Aggregation (DLA)
clusters on the square lattice by extensive numerical simulations
(with clusters having up to $10^8$ particles).  We observe that DLA
clusters undergo strongly anisotropic growth, with the maximal growth
rate along the axes.  The naive scaling limit of a DLA cluster by its
diameter is thus deterministic and one-dimensional.  At the same time,
on all scales from the particle size to the size of the entire cluster
it has non-trivial box-counting fractal dimension which corresponds to
the overall growth rate which, in turn, is smaller than the growth
rate along the axes.  This suggests that the fractal nature of the
lattice DLA should be understood in terms of fluctuations around
one-dimensional backbone of the cluster.
\end{abstract}



% 05.40.-a 	Fluctuation phenomena, random processes, noise, and Brownian motion
% 05.50.+q 	Lattice theory and statistics (Ising, Potts, etc.)
% 05.10.Ln 	Monte Carlo methods
% 05.65.+b 	Self-organized systems

\maketitle

\section{Introduction}

Diffusion Limited Aggregation (DLA) was first introduced by Witten and
Sander \cite{Witten81,Witten83} as a model of irreversible colloidal
aggregation and then rapidly became a basic model of non-equilibrium
growth phenomena such as electrodeposition and dendritic growth,
viscous fingering in fluids, dielectric breakdown, mineral deposition,
bacterial colony growth, pattern formation, to name but a few
\cite{Niemeyer84,Brady84,Matsushita84,Nittmann85,Vicsek,Meakin,Meakin95,Halsey00,Sander00}.
The growth is driven by a Laplacian field and is modeled by adding
particles, one at a time, to a growing cluster via either a random
walk on a lattice, or Brownian motion.  In spite of these very simple
growth rules, only a few rigorous mathematical results about DLA are
available \cite{Kesten87,Kesten90}.  Most properties of both
on-lattice and off-lattice DLA clusters are known either from
numerical simulations, or from theoretical approximations (see
\cite{Muthukumar83,Tokuyama84,Turkevich85,Ball85,Ball86,Family87,Halsey94,Halsey97}
and references therein).  In particular, numerical simulations have
revealed that DLA clusters on the square lattice are inhomogeneous
\cite{Meakin85b,Halsey85,Meakin87}, anisotropic
\cite{Meakin85,Meakin86,Ball85,Ball85b,Kertesz86} and multifractal
\cite{Halsey86,Mandelbrot90}; their properties are lattice dependent
(i.e., nonuniversal) \cite{Meakin86}; their scaling is not determined
by a single exponent \cite{Plischke84,Meakin87}; and the involved
``exponents'' change with the number of particles suggesting a
transient regime \cite{Meakin86,Meakin87}.  To some extent, all these
properties are caused by the local anisotropy of the lattice growth
rules.  As a consequence, the scaling limit of the on-lattice DLA
remains controversial.  This situation contrasts with the significant
progress made over the last decade in the analysis of other lattice
models such as percolation and Ising models.  The identification of
stochastic Loewner evolution (SLE) processes as the scaling limit of
lattice models led to numerous breakthrough discoveries in this field
of statistical physics and mathematics
\cite{Duplantier99,Smirnov01,Chelkak12}.

In this paper, we provide theoretical arguments and extensive
numerical simulations to shed a light onto the scaling limit of
on-lattice DLA clusters.  Our main conclusion is that the naive but
widely used scaling limit, in which the cluster is rescaled by its
diameter, is a deterministic one-dimensional cross-like shape.
Figuratively speaking, anisotropy of the cluster beats fractality,
resulting in a trivial, non-fractal limit.  To explain this point, let
us consider a graph of a one-dimensional random walk (with unit-size
steps) versus the number of steps $t$.  This is a random curve on the
plane.  Rescaling of this curve by its diameter (which is equal to $t$
in this setting) yields a trivial deterministic limit -- the unit
interval.  In order to obtain a nontrivial limit (the Brownian path),
anisotropic rescaling has to be performed, by $t$ and $\sqrt{t}$,
along the horizontal and vertical axes, respectively.  While the
choice of rescaling factors is elementary for this toy example, the
proper rescaling of an anisotropic on-lattice DLA cluster remains
unknown.

The scaling properties of DLA clusters are usually characterized by
two observables: the growth rate $\beta$ and the fractal dimension
$D$.  Most authors compute the latter using the former.  Indeed, if one
covers a {\it regular} fractal of diameter $1$ by disks of size
$\epsilon$, then the number of disks scales as $N \propto
\epsilon^{-D}$, where $D$ is the Minkowski dimension of the fractal.
Rescaling the fractal by $\epsilon^{-1}$ yields the diameter $\propto
N^D$.  Hence the growth rate is the inverse of the dimension.  This
relation that was first put forward by Stanley for percolation
clusters \cite{Stanley77} (see also \cite{Mandelbrot}), was often used
to get the fractal dimension of both on-lattice and off-lattice DLA
clusters by computing the growth rate for the radius of gyration
(e.g., \cite{Meakin86}).  It is important to stress, however, that
this relation does not hold in general, it is valid only under some
regularity assumptions.  The simplest counter-example is an aggregate
of $2t$ particles, half of them forming a disk of radius $\propto
\sqrt{t}$, and the other half forming an interval of length $\propto
t$.  For this aggregate the fractal dimension is $2$ (determined by
the disk) but the growth rate is $1$ (determined by the interval).
The naive rescaling by the diameter $\propto t$ results in a trivial
limit (the unit interval) because the part with the higher dimension
but smaller growth rate (the disk) is shrunk and thus fully eliminated
in the limit $t\to\infty$.

\begin{figure}
\begin{center}
\includegraphics[width=85mm]{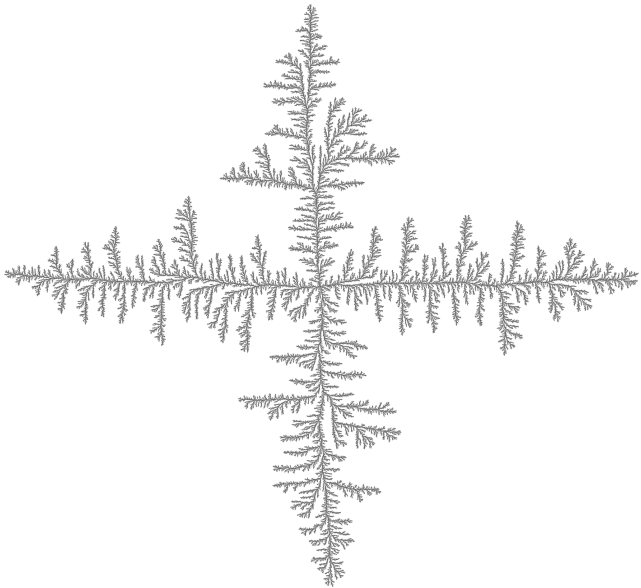}  % {dla18_2048x2048a.eps}
\end{center}
\caption{
A large DLA cluster with with 145~199~976 particles.  In this
coarse-grained picture of resolution $2048\times 2048$, each pixel
represents a $64 \times 64$ block of the original cluster image of
size $2^{17} \times 2^{17}$. }
\label{fig:dla_growth2}
\end{figure}

To our knowledge, the above regularity assumption and the consequent
equality between the inverse of the growth rate $\beta$ and the
fractal dimension $D$ were never properly verified for the on-lattice
DLA.  The first goal of our work is to check this important equality.
Although we obtain slightly different numerical values for $D$ and
$1/\beta$, they cannot be distinguished within the numerical accuracy.
The second goal consists in emphasizing the role of anisotropy.  For
this purpose, we introduce the angular growth rate and show that DLA
clusters grow faster along the axes of the square lattice.  In
particular this implies that if the DLA cluster is rescaled by its
diameter, then the scaling limit becomes deterministic and
one-dimensional.  In other words, the parts of the DLA cluster with
lower growth rates are eliminated, as in the above example with a disk
and an interval.  One can interpret this result as a kind of the law
of large numbers for DLA clusters.  On the other hand, branches of DLA
exhibit a strong pre-fractal behavior that suggests that fluctuations
of DLA branches around the axes have non-trivial scaling limit.  This
observation can be interpreted as an analogue of the central limit
theorem.

\section{Numerical results}

Our strategy to support the above claims consists in two parts: (i)
numerical computation of both the growth rate $\beta$ and the fractal
dimension $D$, and (ii) profound analysis of the cluster anisotropy.
For this purpose, we adapted a bias-free algorithm by Y. E. Loh to
generate DLA clusters on the square lattice \cite{Loh14}.  The growth
of each cluster was stopped when it reaches the edges of the square
computational domain, $2^{\lmax} \times 2^{\lmax}$, with a prescribed
scale $\lmax$.  As a result, the number of particles in various
clusters is not identical.  We generated 100 clusters with $\lmax =
16$ that have the minimal and the maximal number of particles
41~003~402 and 51~514~999, respectively.  We also generated one larger
cluster with 145~199~976 particles by setting $\lmax = 17$
(Fig. \ref{fig:dla_growth2}).  To our knowledge, this is the largest
{\it on-lattice} DLA cluster ever generated (in contrast, off-lattice
DLAs of similar sizes have been earlier reported, e.g.,
\cite{Mandelbrot02}).

\subsection{Fractal dimension versus growth rate}

\begin{figure}
\begin{center}
\includegraphics[width=80mm]{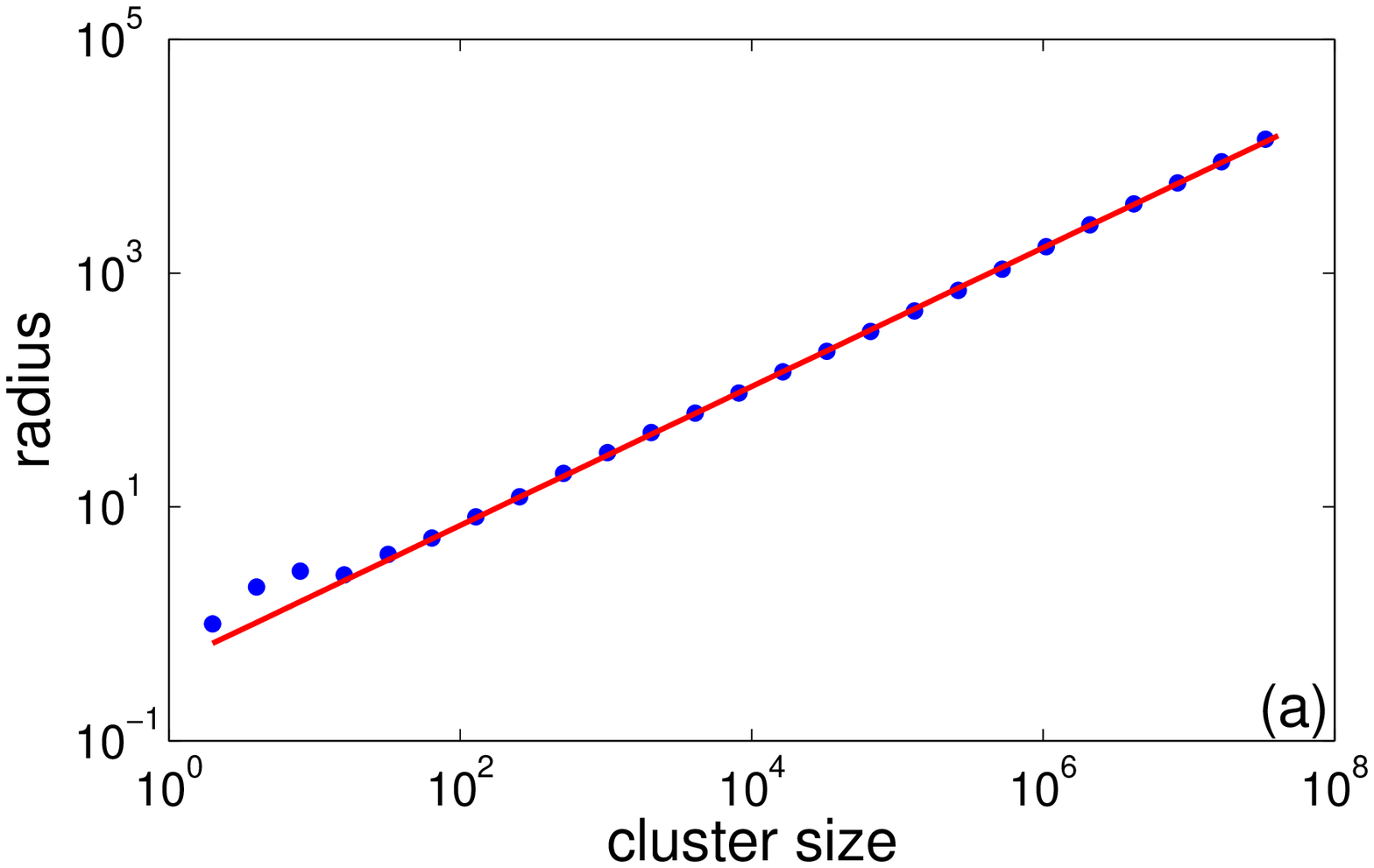}  % {rg001.eps}
\includegraphics[width=80mm]{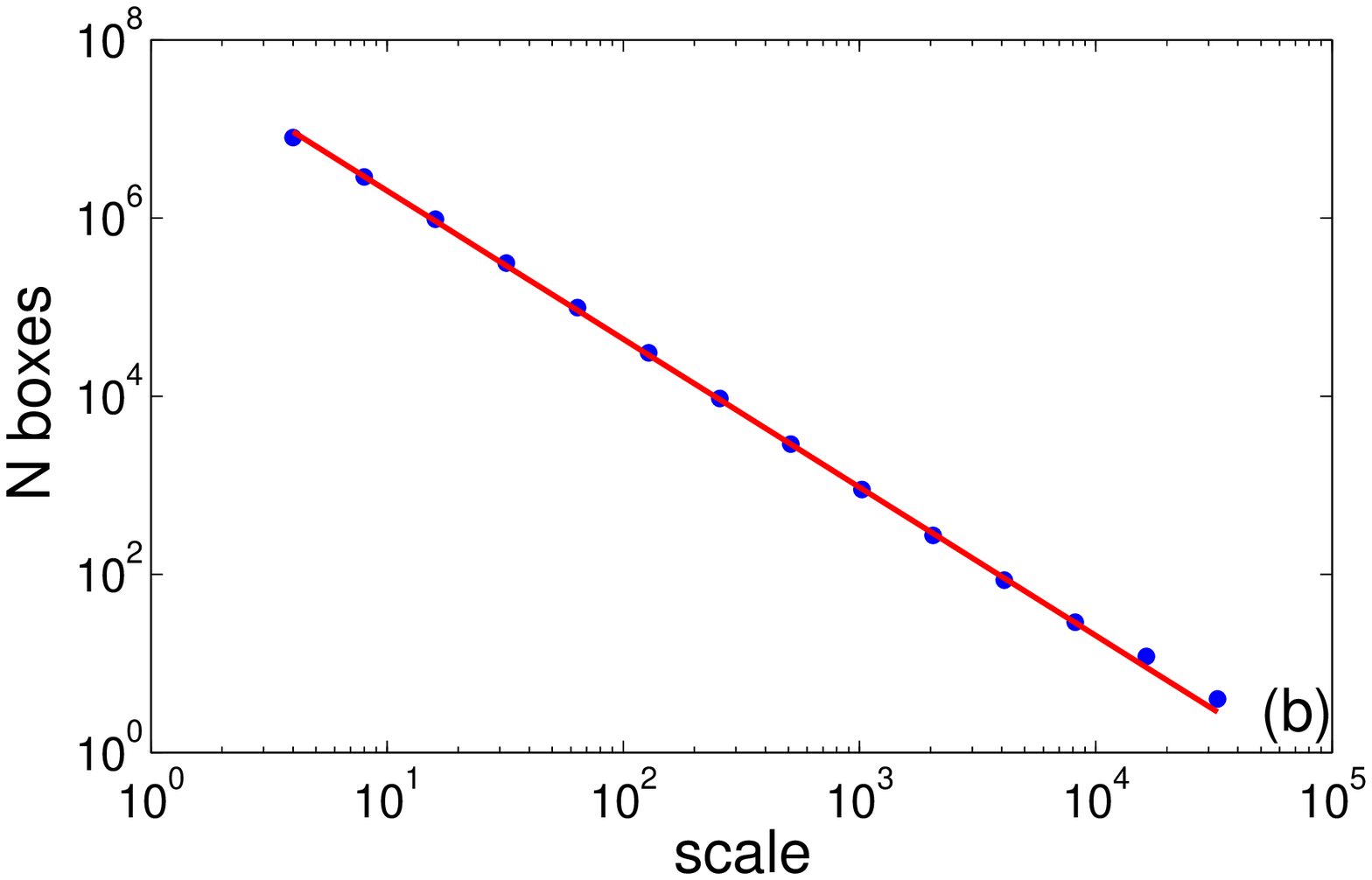}  % {Dbox001.eps}
\end{center}
\caption{
(Color online) {\bf (a)} Radius of gyration $R(t)$ as a function of
the cluster size $t$ for one cluster (symbols).  A linear fit at
loglog scale (line), $\ln R(t) = 0.594 \ln t -0.796$, was obtained
over sizes from $2^{10}$ to $2^{25}$.  {\bf (b)} Number of non-empty
boxes, $\N_\ell$, at scale $\ell$, for the same cluster.  Solid line
shows a linear fit at loglog scale, $\ln \N_\ell = -1.664 \ln \ell +
18.351$, obtained for scales ranging between $2^2$ and $2^{13}$.  }
\label{fig:rg}
% A_DLA_analysis(1, dla,B);   % attention, it takes 7 minutes to compute B
\end{figure}

Knowing the history of growth of each generated cluster, we compute
two conventional characteristics: the cluster radius, ${\mathcal
R}(t)$, and the radius of gyration, $R(t)$, as functions of the
cluster size $t$ (i.e., the number of particles)
\begin{eqnarray}
\label{eq:Rmax}
{\mathcal R}(t) &=& \max_{1\leq k \leq t} \left\{ \sqrt{x_k^2 + y_k^2} \right\} , \\
R(t) &=& \left(\frac{1}{t} \sum\limits_{k=1}^t  (x_k^2 + y_k^2)\right)^{\frac12} , 
\end{eqnarray}
where $(x_k,y_k)$ are the coordinates of the $k$-th attached particle
(with the seed point of the cluster, $(x_0,y_0)$, being located at the
origin).  We checked that ${\mathcal R}(t)$ and $R(t)$ behave
similarly, being different by a factor between 2 and 3.  For this
reason, we focus on the radius of gyration which exhibits less
fluctuations.  Figure \ref{fig:rg}a illustrates a power law growth of
$R(t)$ with the cluster size $t$ for one DLA cluster.  For this
cluster, we get the growth exponent $\beta = 0.594$.  This value is by
$2\%$ larger than the earlier reported growth exponent of off-lattice
DLA, $0.583$ \cite{Tolman89}.  For the same cluster, we also compute
its box-counting dimension by evaluating the number $\N_\ell$ of
non-empty boxes at scale $\ell$, ranging from $1$ (the size of one
particle) to $2^{\lmax}$ (the size of the whole cluster).  Figure
\ref{fig:rg}b shows a power law scaling $\N_\ell \propto \ell^{-D}$,
which enables us to determine the Minkowski (box-counting) dimension
$D$.  For this DLA cluster, we get $D \approx 1.664$.  We conclude
that the fractal dimension is smaller than $1/\beta \approx 1.684$ by
$1\%$.  In order to check the relevance of this difference, we
repeated the above analysis for 100 independently generated clusters.
%Performing linear fits at loglog scales, we get the averaged exponents
%and their standard deviations summarized in Table \ref{tab:exponents}.
We obtain the empirical mean and standard deviation for two exponents:
\begin{equation}
D = 1.666 \pm 0.004, \qquad \beta = 0.596 \pm 0.004 .
\end{equation}
The average fractal dimension $D$ is smaller than the average of the
inverse of the growth rate, $1/\beta = 1.678 \pm 0.011$, by only
$0.7\%$, and this difference is below the statistical uncertainty.
For this reason, we cannot exclude the relation $D = 1/\beta$ for
on-lattice DLA clusters.  We also found that the box-counting fractal
dimension $D$ is remarkably close to the theoretical value $5/3$
predicted by mean field theories \cite{Muthukumar83,Tokuyama84}, an
analytical diamond-shaped model \cite{Ball85}, and a continuous-time
random walk theory \cite{Turkevich85}.

%\begin{table}
%\begin{center}
%\begin{tabular}{| c | l | c | c |} \hline
%        & Scaling                                          & Exponent          & Dimension         \\  \hline
%$D$     & $N_\ell \propto \ell^{-D}$                         &                   & $1.666 \pm 0.004$ \\ 
%$\beta$ & $R(t) \propto t^\beta$                         & $0.596 \pm 0.004$ & $1.678 \pm 0.011$ \\  
%$\beta_{\rm axis}$ & $R_{\rm axis}(t) \propto t^{\beta_{\rm axis}}$ & $0.631 \pm 0.014$ & $1.586 \pm 0.035$ \\
%$\beta_{\rm diag}$ & $R_{\rm diag}(t) \propto t^{\beta_{\rm diag}}$ & $0.535 \pm 0.021$ & $1.873 \pm 0.073$ \\  \hline
%\end{tabular}
%\end{center}
%\caption{
%The box-counting dimension $D$ and the growth rates $\beta$,
%$\beta_{\rm axis}$, $\beta_{\rm diag}$ (and the associated
%``dimensions'' $1/\beta$, $1/\beta_{\rm axis}$, $1/\beta_{\rm diag}$)
%obtained from linear fits at loglog scale for each DLA cluster with
%$\lmax = 16$.  The mean and standard deviation are obtained by
%averaging over $100$ independent clusters.  For the computation of the
%angular radius of gyration, we used $n_s = 32$.}
%\label{tab:exponents}
%% [Rgall,Rgaxis,Rgdiag,Rgyrall,Dbox] = A_DLA_analysis_mean();
%\end{table}

% Latest computation: 02/08/2017
%D-Rall : 1.677626e+000 +/- 1.136938e-002
%D-Raxis : 1.585811e+000 +/- 3.490364e-002
%D-Rdiag : 1.874892e+000 +/- 7.187930e-002
%beta-Rall : 5.961075e-001 +/- 4.037308e-003
%beta-Raxis : 6.308971e-001 +/- 1.400953e-002
%beta-Rdiag : 5.341314e-001 +/- 2.025101e-002
%Dbox : -1.666410e+000 +/- 3.711153e-003

\subsection{Role of anisotropy}

Various measures have been introduced in 1980's to characterize the
anisotropy of DLA clusters on the square lattice
\cite{Meakin85,Meakin86,Ball85,Ball85b,Kertesz86}.  We propose another
quantity, the {\it angular growth rate}, which is much better adapted
to the study of anisotropic but mostly star-like structures.  We cover
the plane by $n_s$ equal sectors $S_1,\ldots, S_{n_s}$ (of angle
$2\pi/n_s$) centered at the origin (the center of the cluster), and
define the angular radius of gyration up to cluster size $t$:
\begin{equation}
R_\theta(t) = \left(\frac{1}{n_\theta(t)} \sum\limits_{k=1}^t  (x_k^2 + y_k^2) \,  {\mathcal I}_{(x_k,y_k) \in S_\theta} \right)^{\frac12} ,
\end{equation}
where ${\mathcal I}_{(x_k,y_k) \in S_\theta}$ is equal $1$ if the
point $(x_k,y_k)$ belongs to the sector $S_\theta$ of a discretized
polar angle $\theta$, and zero otherwise, while $n_\theta(t)$ is the
number of cluster points belonging the sector $S_\theta$ up to $t$.
The angular growth rate, $\beta_\theta$, is defined from the expected
power law scaling: $R_\theta(t) \propto t^{\beta_\theta}$ as
$t\to\infty$.  In this way, one can probe whether the growth rate
depends on the direction and, in particular, whether the growth rates
along the square lattice axes and along the diagonals are different.

\begin{figure}
\begin{center}
\subfigure[~$t = 10^4$]{\includegraphics[width=40mm]{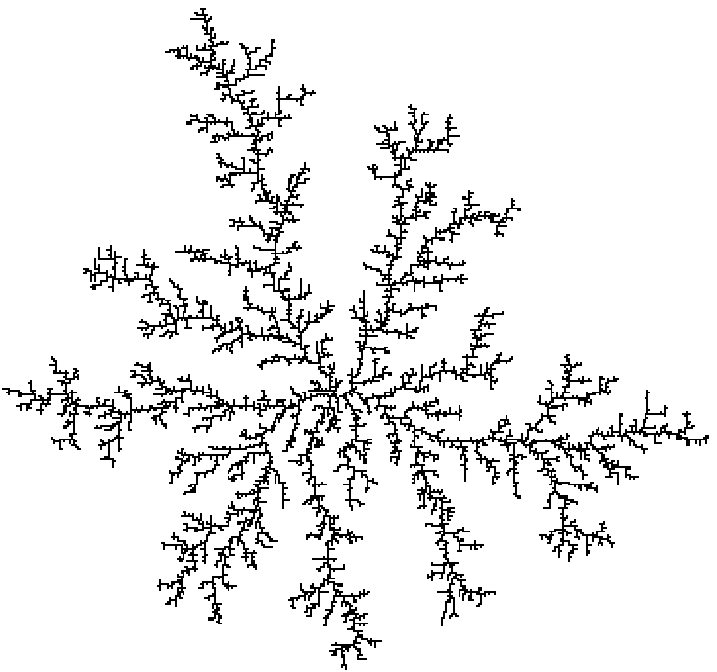}} % {dla18_001_coarse_9_10000a.eps}
\subfigure[~$t = 10^4$]{\includegraphics[width=40mm]{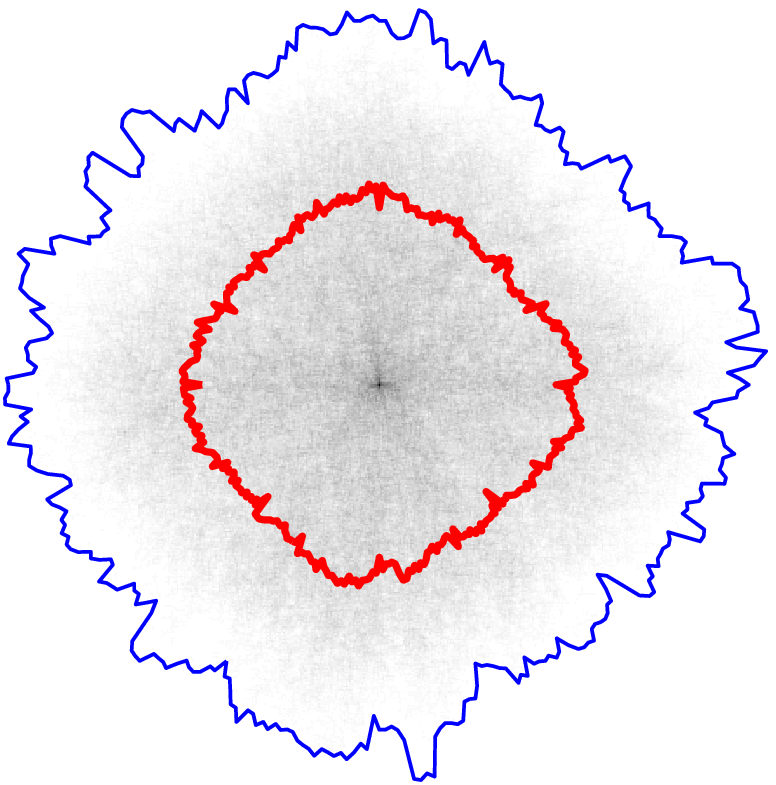}} % {sumdla4b.eps}
\subfigure[~$t = 10^5$]{\includegraphics[width=40mm]{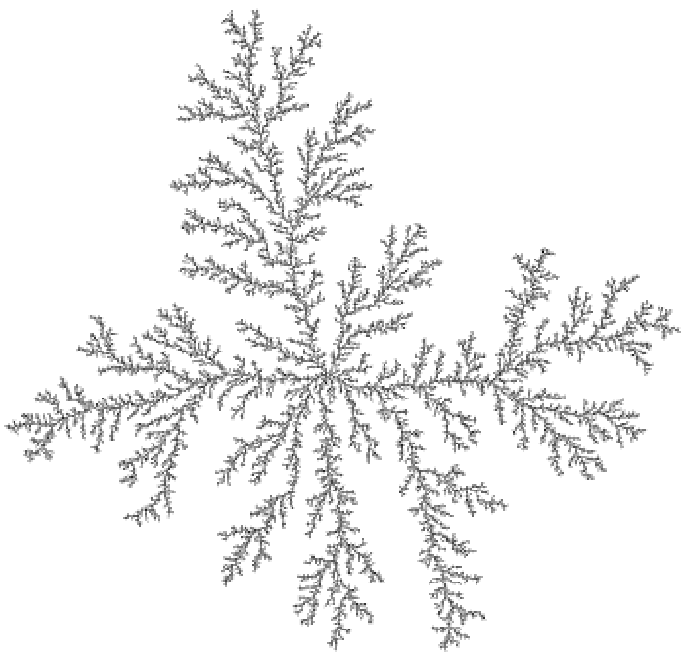}} % {dla18_001_coarse_11_100000a.eps}
\subfigure[~$t = 10^5$]{\includegraphics[width=40mm]{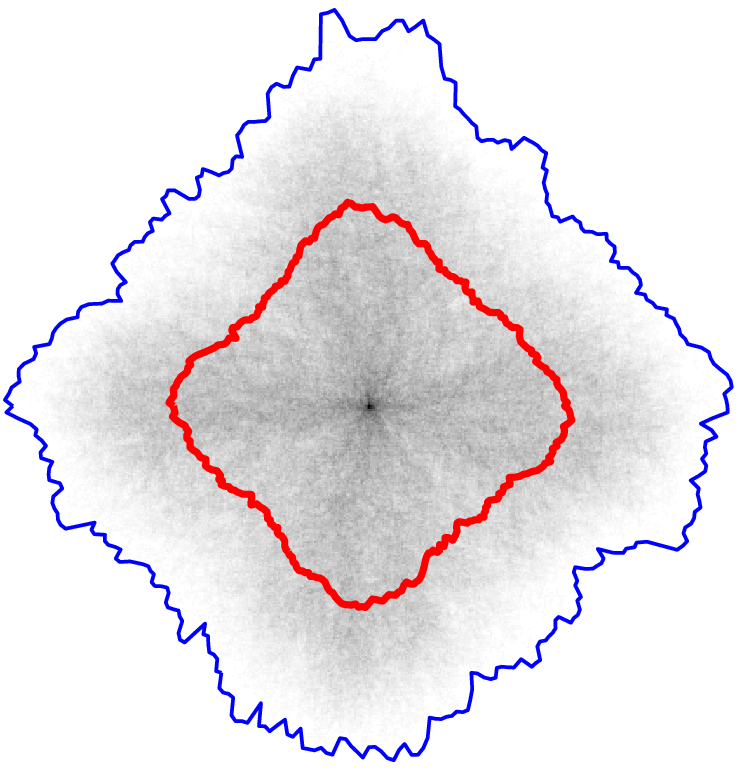}} % {sumdla5b.eps}
\subfigure[~$t = 10^6$]{\includegraphics[width=40mm]{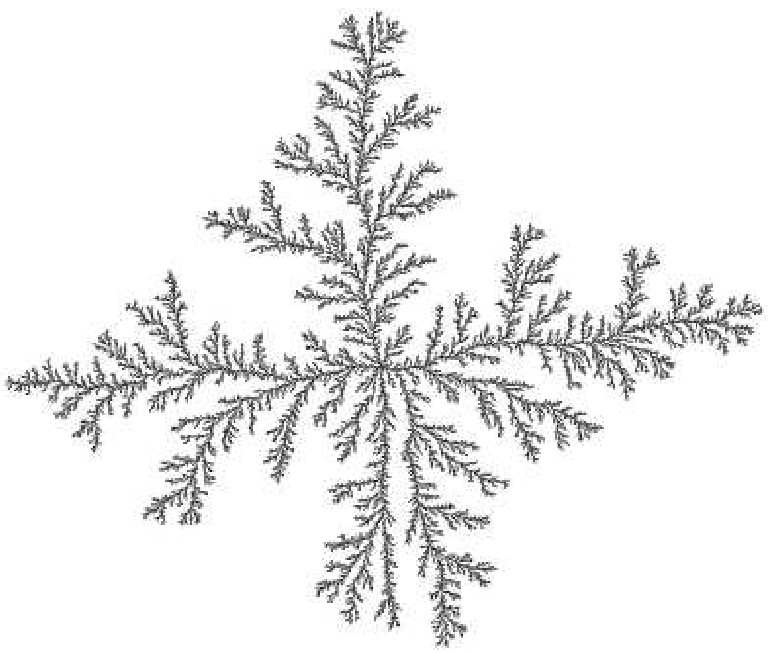}} % {dla18_001_coarse_13_1000000a.eps}
\subfigure[~$t = 10^6$]{\includegraphics[width=40mm]{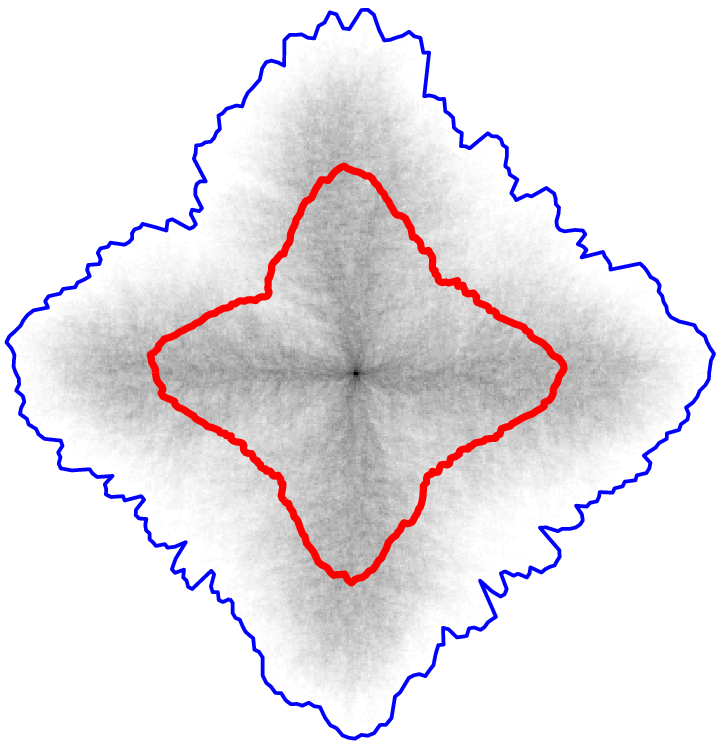}} % {sumdla6b.eps}
\subfigure[~$t = 10^7$]{\includegraphics[width=40mm]{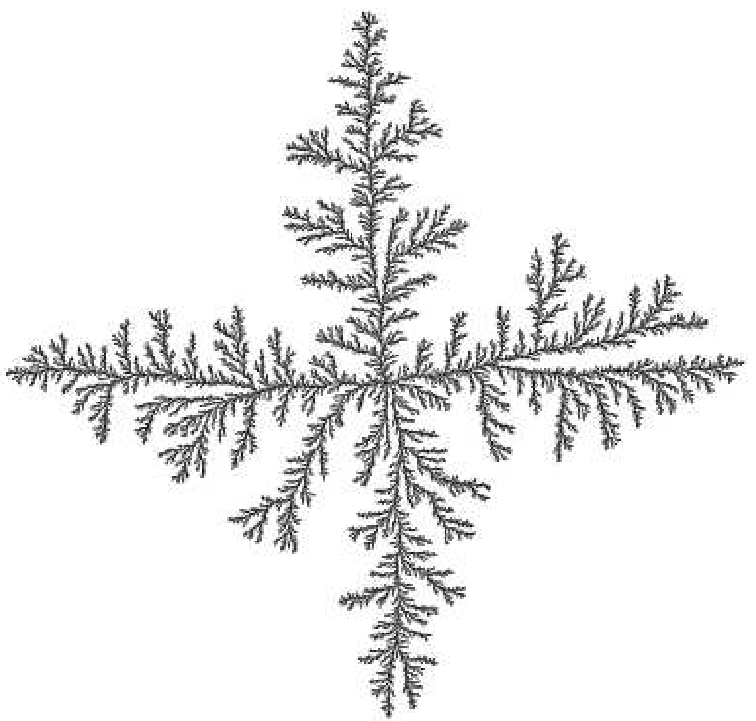}} % {dla18_001_coarse_15_10000000a.eps}
\subfigure[~$t = 10^7$]{\includegraphics[width=40mm]{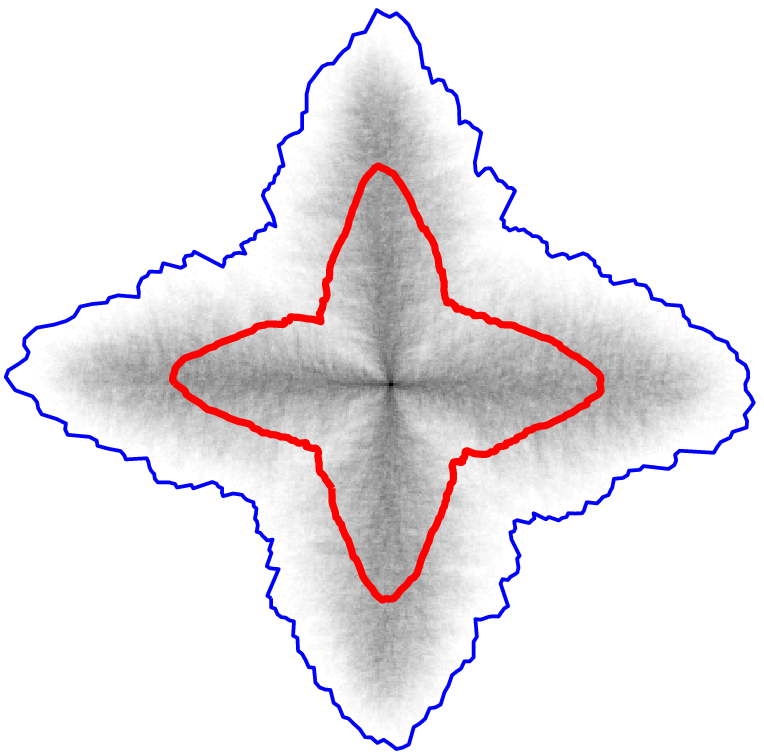}} % {sumdla7b.eps}
\end{center}
%\begin{picture}(0,0)(0,0)
%\put(-120,465){$t = 10^4$}
%\put(-120,345){$t = 10^5$}
%\put(-120,225){$t = 10^6$}
%\put(-120,115){$t = 10^7$}
%\end{picture}
\caption{
(Color online).  {\bf Left column:} A DLA cluster at various cluster
sizes $t$: $10^4$, $10^5$, $10^6$, and $10^7$, from the top to the
bottom (coarse-grained $512\times 512$ pictures); {\bf Right column:}
Grayscale representation of the density of points, averaged over 100
DLA clusters, at the same $t$; thin outer contour shows the maximally
distant points from the center (an angular version of the maximal
distance ${\mathcal R}$ defined by Eq. (\ref{eq:Rmax})); thick inner
contour shows the angular radius of gyration, with the angular
resolution of $1^\circ$.}
\label{fig:dla_growth}
% A_DLA_meancluster_plot(7);
\end{figure}

The left column of Fig. \ref{fig:dla_growth} shows the progressive
growth of the largest DLA cluster shown in Fig. \ref{fig:dla_growth2}.
One can clearly see how an isotropic structure of a small cluster
(with $10^4$ particles) slowly evolves into the cross-like anisotropic
structure of larger clusters (e.g., with $10^7$ particles).  For
comparison, the right column of Fig. \ref{fig:dla_growth} shows a
grayscale representation of the density of points, averaged over 100
DLA clusters, at the same $t$.  The average density is defined as the
sum of indicator functions of 100 independently generated clusters.
For small clusters ($t = 10^4$ and below), the density is almost
isotropic, meaning that the typical cluster has almost a round shape.
For larger clusters with $t = 10^5$, the diamond shape emerges,
indicating a directional preferential growth along the four axes.  At
$t = 10^6$ and $t = 10^7$, the diamond shape progressively transforms
into a cross-like shape.  These features are particularly well seen by
looking at two contours: the outer contour showing the maximally
distant points from the center, and the inner contour showing the
angular radius of gyration.  These two contours were computed by
identifying the points of all 100 clusters that lie within a sector
between angles $\theta$ and $\theta + \delta\theta$ (with the angular
resolution $\delta\theta = 1^\circ$, i.e., $n_s = 360$).  In each
sector, the distance between the center and the most distant point,
and the angular radius of gyration $R_\theta(t)$, were computed and
then plotted versus the polar angle $\theta$ from $0$ to $360^\circ$.
The outer and inner contours illustrate respectively the positions of
extreme and average points of DLA clusters.  Remarkably, these two
contours evolve with the cluster size in a very similar way.  Note
that the evolution of a commonly observed diamond-like structure of
the square DLA clusters into a cross-like shape with four distinct
arms was first conjectured by Meakin \cite{Meakin86} and then
confirmed by numerical simulations on larger clusters in
\cite{Meakin87}.  Moreover, the scaling exponents for the length and
width of the four main arms were claimed to be different
\cite{Turkevich85,Family87,Meakin87}.  We note however that the
arguments elaborated in these papers rely on (over)simplified
assumptions (e.g., the diamond-like limiting shape of DLA clusters),
whereas predictions of the scaling exponents were sometimes different.
While there was not doubt about anisotropic character of the
on-lattice DLA growth, its explanations remained rather controversial.

\begin{figure}
\begin{center}
\includegraphics[width=80mm]{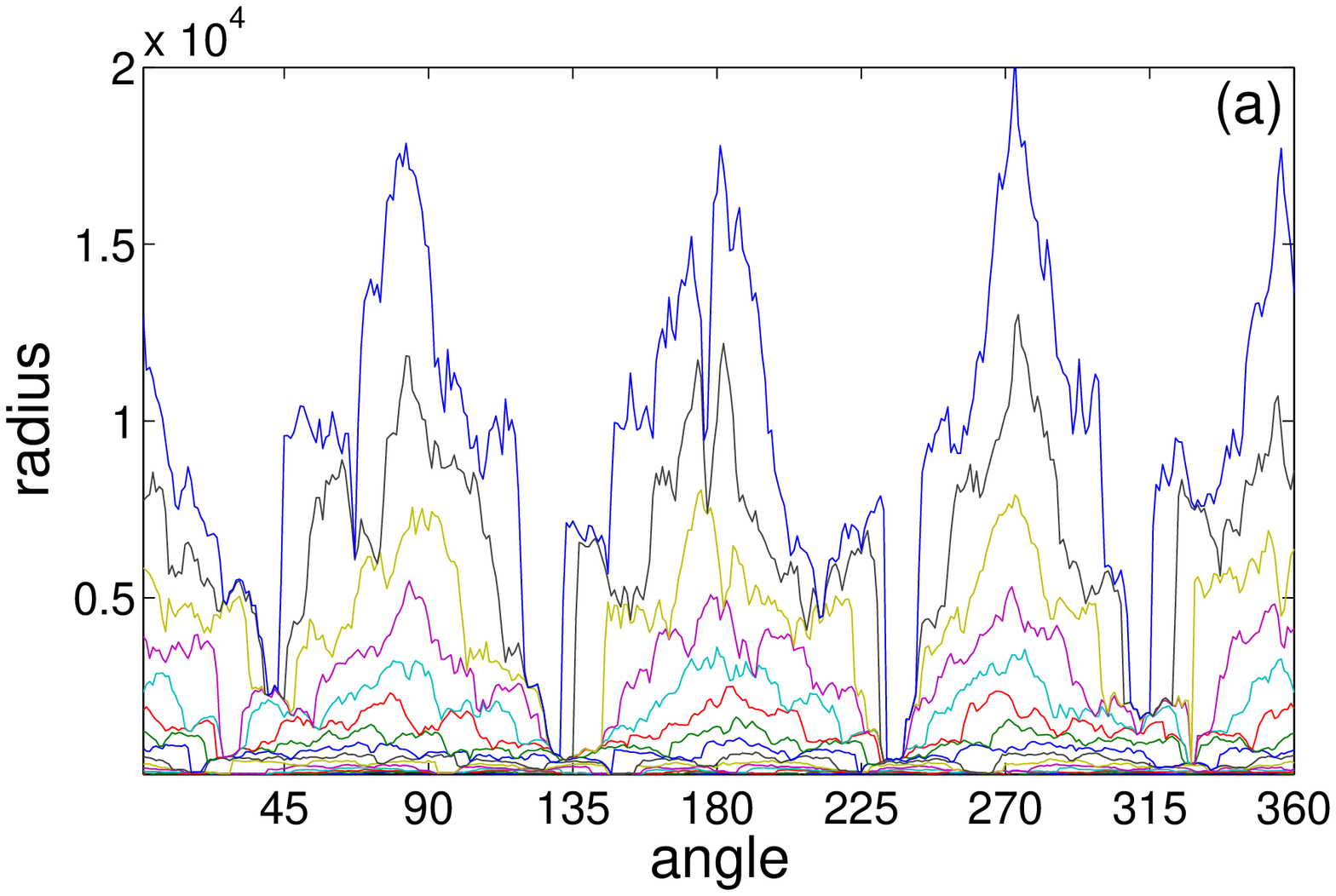} % {rg_alpha001_360.eps}
\includegraphics[width=80mm]{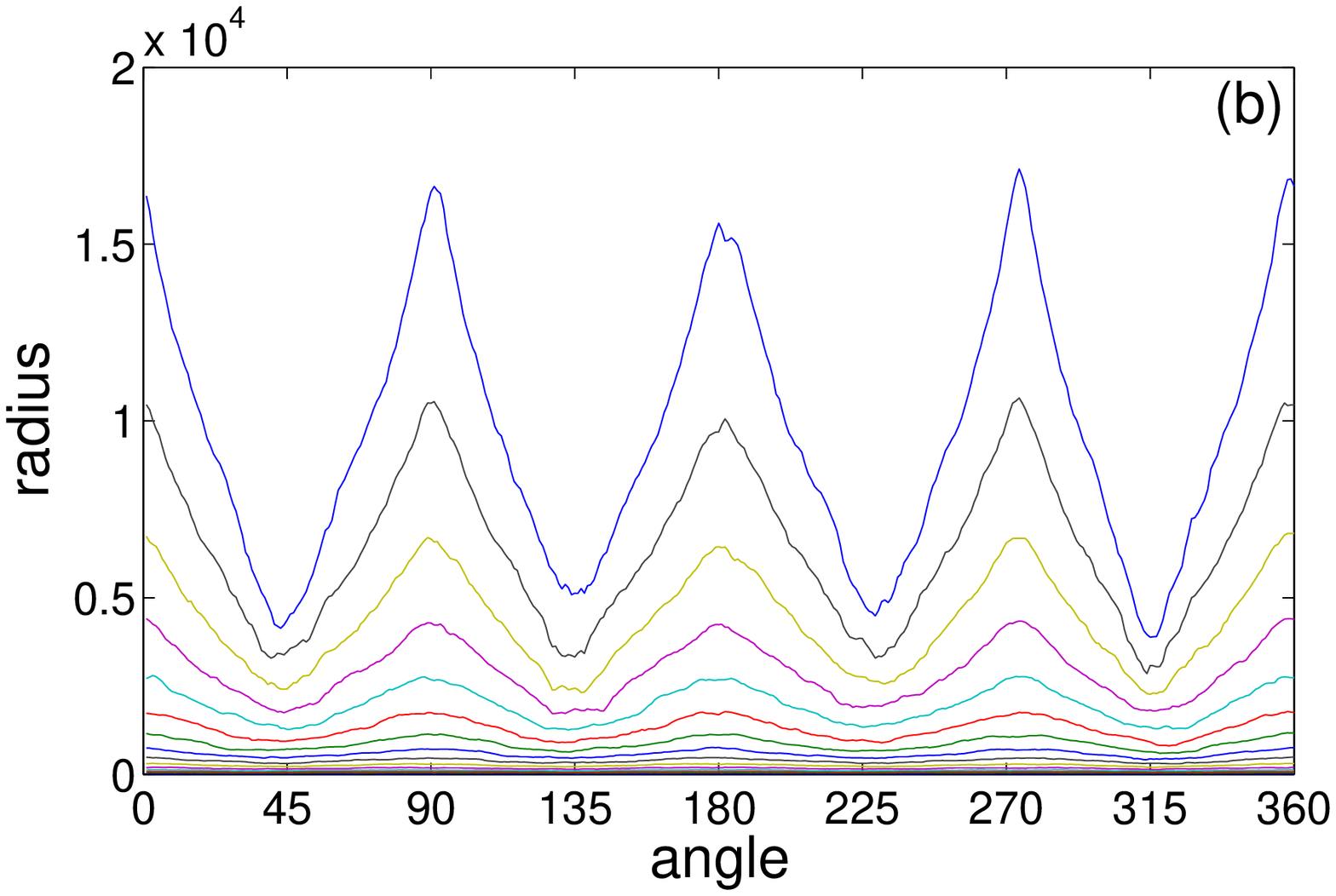} % {rg_alpha_mean_360.eps}
\includegraphics[width=80mm]{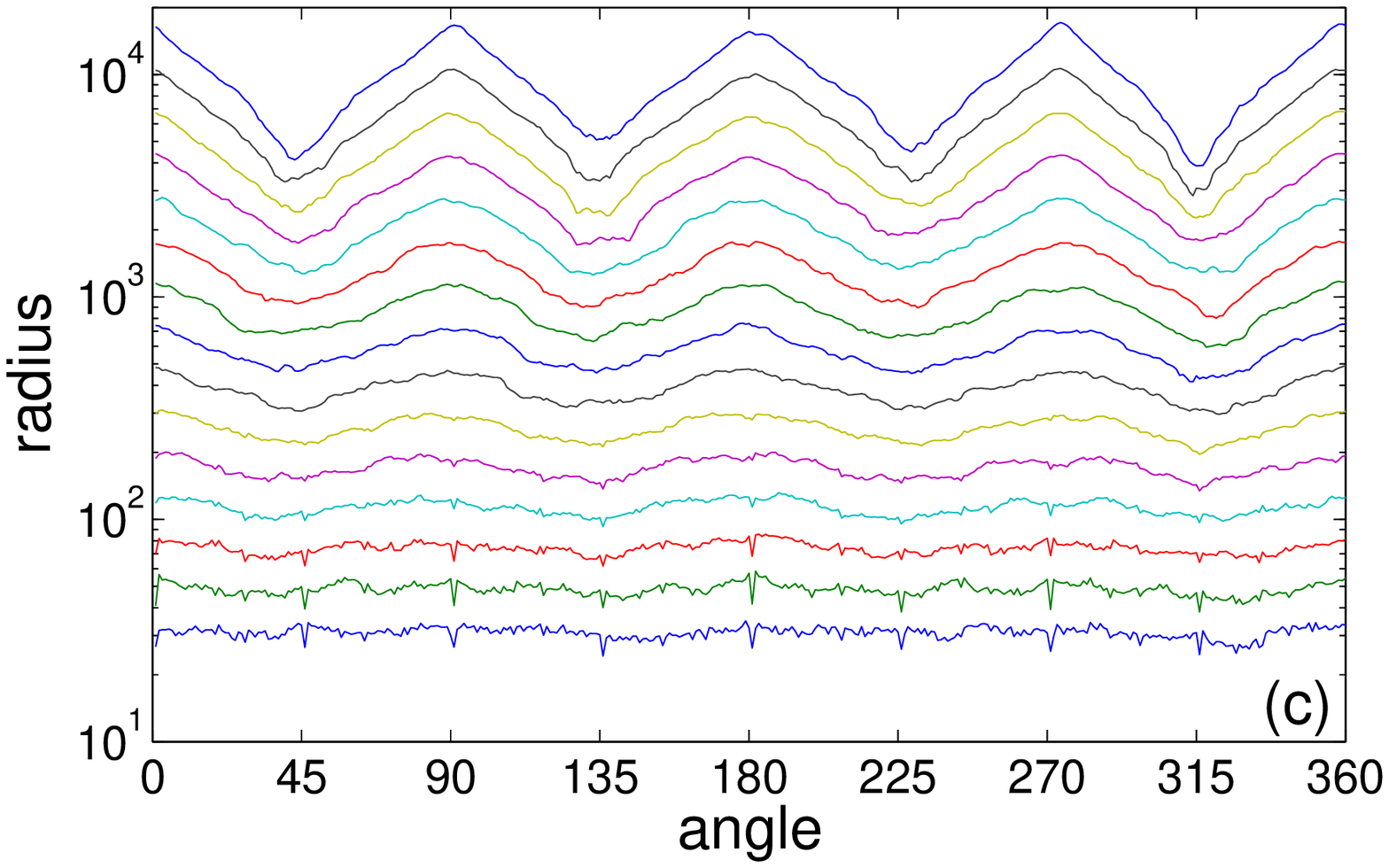} % {rg_alpha_mean_360_log.eps}
\end{center}
\caption{
(Color online) {\bf (a)} Angular radius of gyration $R_\theta(t)$ as a
function of the angle $\theta$ for one cluster, at cluster sizes $t =
2^{11}, 2^{12}, \ldots, 2^{25}$ (corresponding curves are arranged
from bottom to top).  {\bf (b,c)} Angular radius of gyration
$R_\theta(t)$ as a function of the angle $\theta$, averaged over 100
clusters, at cluster sizes $t = 2^{11}, 2^{12}, \ldots, 2^{25}$, at
linear {\bf (b)} and semilogarithmic {\bf (c)} scales.  For all plots,
we set $n_s = 360$. }
\label{fig:R_alpha}
% A_DLA_analysis(1, dla,B);
% [Rgall,Rgaxis,Rgdiag,Rgyrall,Dbox] = A_DLA_analysis_mean();
\end{figure}

After this visual inspection, we proceed to quantify the anisotropic
effects.  Figure \ref{fig:R_alpha}a shows how the angular radius of
gyration depends on the direction $\theta$ at different cluster sizes
$t$.  One can see how the anisotropy is progressively established
(with four maxima along axes and four minima along diagonal
directions).  For comparison, Fig. \ref{fig:R_alpha}b shows the
angular radius of gyration $R_\theta(t)$ for an average cluster
obtained by superimposing 100 clusters.  As expected, this plot
resembles that for one cluster but the average over 100 clusters
yields smoother curves.  The emergence of anisotropy is particularly
clear at semilogarithmic scale (Fig. \ref{fig:R_alpha}c): flat
profiles of $R_\theta(t)$ versus $\theta$ at small cluster sizes $t$
progressively become uneven, with prominent peaks in four axial
directions.

\begin{figure}
\begin{center}
\includegraphics[width=80mm]{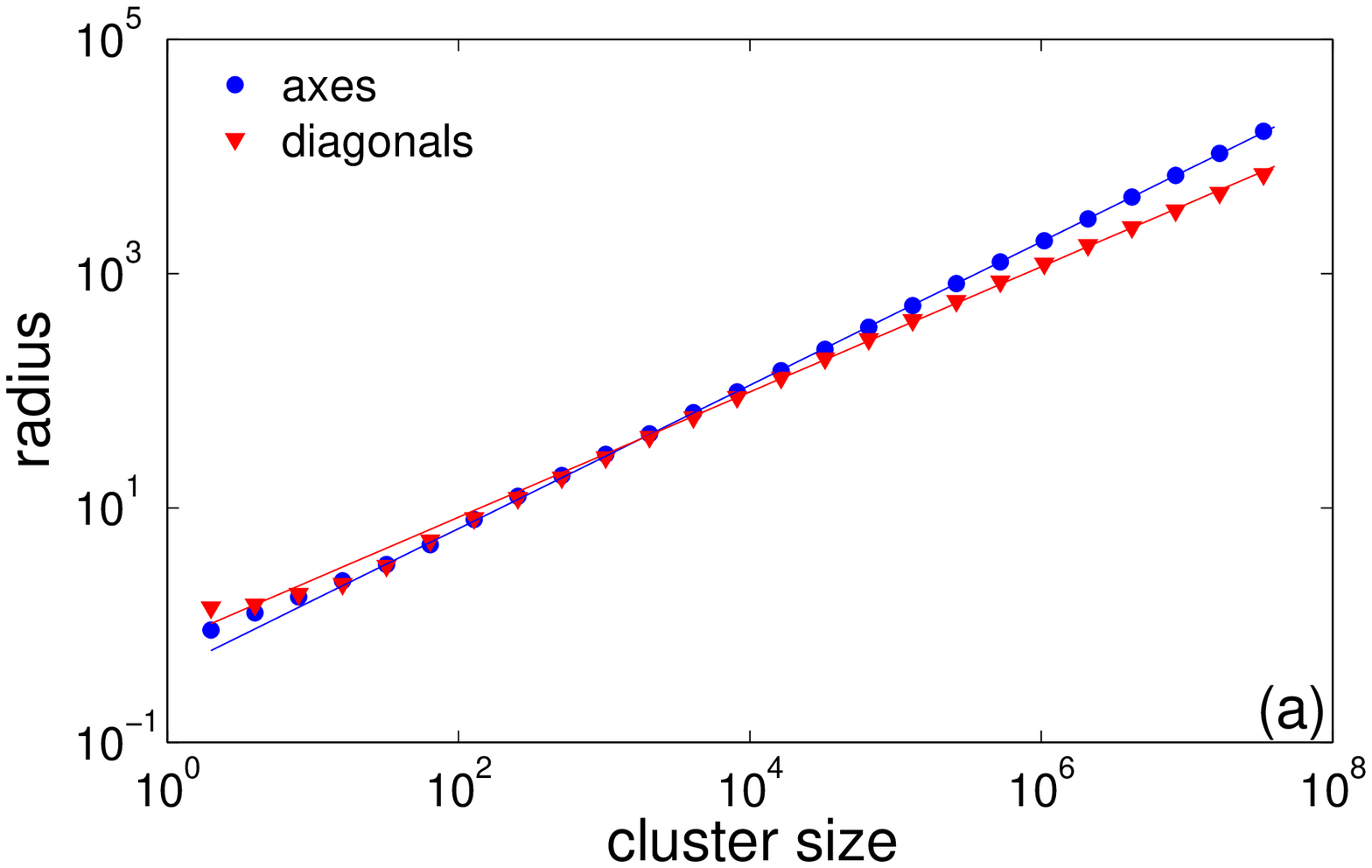} % {rg_axis_mean.eps}
\includegraphics[width=80mm]{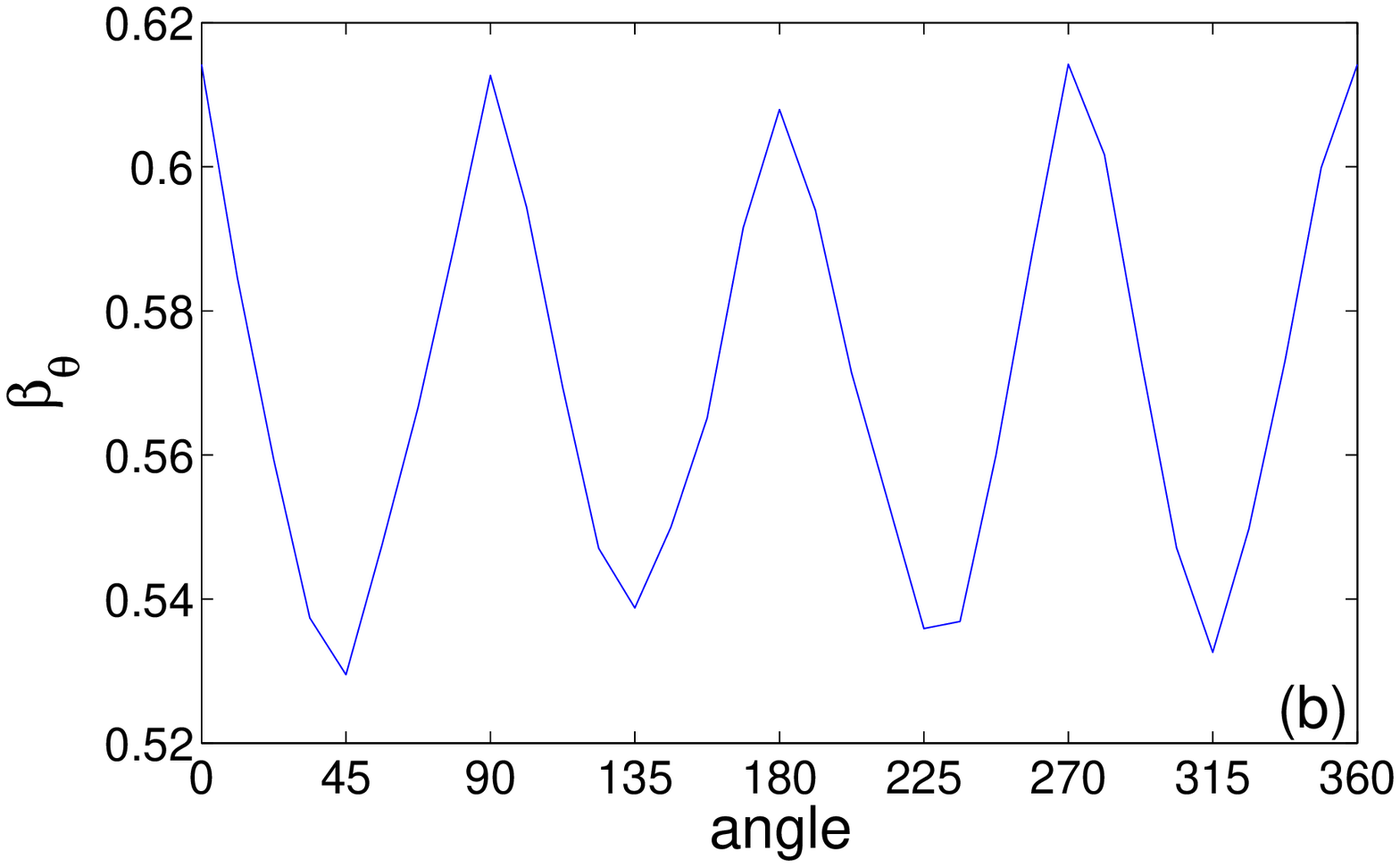} % {beta_theta.eps}
\end{center}
\caption{
(Color online) {\bf (a)} Aggregated angular radii of gyration for 4
axes, $R_{\rm axis}(t)$, and for 4 diagonals, $R_{\rm diag}(t)$,
averaged over 100 DLA clusters, at cluster sizes $t = 1, 2, 2^2,
\ldots, 2^{25}$.  Linear fits at loglog scale
(lines) are: $\ln R_{\rm axis}(t) = 0.612 \ln t - 0.923$ and $\ln
R_{\rm diag}(t) = 0.535 \ln t - 0.340$.  {\bf (b)} The angular growth
rate $\beta_\theta$ as a function of the angle $\theta$, obtained from
linear fits at loglog scale of $R_\theta(t)$ versus $t$.  For both
plots, we set $n_s = 32$.  }
\label{fig:R_axis}
% A_DLA_analysis(1, dla,B);
\end{figure}

In order to reveal the different growth along axes and diagonals, we
aggregate the angular radii of gyration $R_\theta(t)$ for 4 directions
of square lattice axes to define $R_{\rm axis}(t)$, and for 4 diagonal
directions to define $R_{\rm diag}(t)$.  
%
%Fitting these radii at the loglog scale, we obtain the growth rates
%$\beta_{\rm axis}$ and $\beta_{\rm diag}$ for each cluster.  Their
%mean values and standard deviations are summarized in Table
%\ref{tab:exponents}.  
%
Since the number of points in each sector is significantly smaller
than in the whole cluster, fluctuations are much stronger.  For their
reduction, we choose relatively large sectors of angle $11.25^\circ$
(with $n_s = 32$) and we average the aggregated radii over 100 DLA
clusters.  The resulting axial and diagonal radii $R_{\rm axis}(t)$
and $R_{\rm diag}(t)$ are shown in Fig. \ref{fig:R_axis}a.  One can
see the faster growth along the axes than along the diagonals, with
the growth rates $0.612$ and $0.535$, respectively.
%
%While the growth rate along the diagonals is the same as in Table
%\ref{tab:exponents}, the growth rate along the axes is slightly
%smaller.  This difference is not surprising as the average of
%exponents is in general not equivalent to the exponent of an averaged
%cluster.  
%
Finally, Fig. \ref{fig:R_axis}b presents the angular growth rate
$\beta_\theta$ obtained by linear fits at loglog scale of
$R_\theta(t)$ versus $t$ (to reduce fluctuations, the angular radius
of gyration was averaged over 100 DLA clusters).  We observe
variations of $\beta_\theta$ from $0.53$ to $0.61$, the minimal and
maximal growth rates corresponding to the diagonals and to the axes,
respectively.

\section{Discussion}

With the aid of extensive numerical simulations, we have shown that
DLA clusters on the square lattice exhibit strong anisotropic behavior
driven by the local aggregation rules.  In particular, the growth rate
depends on direction, with the maximal growth rate along the axes and
the minimal one along the diagonals.  This implies that after
rescaling by cluster's diameter, the mean size of the cluster in all
non axial directions converges to zero hence the scaling limit becomes
deterministic and one-dimensional.  On the other hand, on all scales
from the particle size to the size of the entire cluster it has
non-trivial box-counting fractal dimension which corresponds to the
overall growth rate of the cluster.  The latter is a sort of a
(non-arithmetic) average of the angular growth rates.  This average
fully ignores distinctions between growth rates in different
directions and thus partly neglects anisotropic effects, in
particular, the fastest growth along the axes.  This suggests that the
fractal nature of the lattice DLA should be understood in terms of
fluctuations around one-dimensional backbone of the cluster.

The crucial impact of the lattice anisotropy on the DLA growth
naturally disappears for the intrinsically isotropic off-lattice DLA
that had also been extensively investigated (see
\cite{Tolman89,Vicsek90,Ossadnik91,Mandelbrot02,Adams09,Hanan12,Menshutin12}
and references therein).  This rises an important question: how to
explain the qualitative difference between the on-lattice and
off-lattice DLA?  There are two obvious distinctions.  First, the
growth of the on-lattice DLA is controlled by the discrete harmonic
measure versus the continuous one for the off-lattice model.  We
believe that this is not really an issue, since the discrete measure
converges rapidly to the continuous one as the size of the aggregate
increases \cite{Komlos76,Kozdron05}.  The second reason is that the
models have different rules for the {\em local} particle attachment.
The effect of the local rules on the anisotropy of the clusters has
been observed before \cite{Bogoyavlenskiy02}.  Given a cluster, the
distribution of places where a new particle will get close to the
cluster is almost the same for particles performing random walk and
for particles performing Brownian motion.  By the concentration of the
harmonic measure, the particle which is started near the cluster will
attach to the cluster very close to its starting position, with a
large probability.  The main difference is that for the lattice there
are very few places where the particle can attach, especially in the
vicinity of a ``tip'' in the cluster.  This has two consequences: (i)
the on-lattice particle is more likely to attach to the tip in the
direction of the fastest growth than the Brownian particle would do,
and (ii) a growing tip is less likely to be split into two competing
branches.  As a consequence, the branches of off-lattice DLA are more
wiggly.

We conclude that the naive but widely used scaling limit of on-lattice
DLA fails due to anisotropy.  The future analysis needs to account for
anisotropic effects and to potentially focus on individual branches of
large DLA clusters.  Our results present thus the first step towards
finding a proper rescaling of DLA clusters that is crucial to
understand the fractal properties of the on-lattice DLA model and its
scaling limit.

\subsection{Acknowledgments}
%\begin{acknowledgments}
DG acknowledges the support under Grant No. ANR-13-JSV5-0006-01 of the
French National Research Agency.  DB acknowledges the support under
EPSRC Fellowship EP/M002896/1.
%\end{acknowledgments}

%%%%%%%%%%%%%%%%%%%%%%%%%%%%%%%%%%%%%%%%%%%%%%%%%%%%%%%%%%%%%%%%%%%%%%%%%%%%%%%%%%%%%%


\begin{thebibliography}{40}


\bibitem{Witten81}		T. A. Witten and L. M. Sander,
				``Diffusion-Limited Aggregation, a Kinetic Critical Phenomenon'',
				Phys. Rev. Lett. {\bf 47}, 1400 (1981).

\bibitem{Witten83}		T. A. Witten and L. M. Sander, 
				``Diffusion-limited aggregation'', 
				Phys. Rev. B {\bf 27}, 5686 (1983).


\bibitem{Niemeyer84}		L. Niemeyer, L. Pietronero, and H. J. Wiesmann,
				``Fractal Dimension of Dielectric Breakdown'',
				Phys. Rev. Lett. {\bf 52}, 1033 (1984).

\bibitem{Brady84}		R. M. Brady and R. C. Ball,
				``Fractal growth of copper electrodeposits'',
				Nature {\bf 309}, 225-229 (1984).

\bibitem{Matsushita84}		M. Matsushita, M. Sano, Y. Hayakawa, H. Honjo, and Y. Sawada,
				``Fractal Structures of Zinc Metal Leaves Grown by Electrodeposition'',
				Phys. Rev. Lett. {\bf 53}, 286 (1984).

\bibitem{Nittmann85}		J. Nittmann, G. Daccord, and H. E. Stanley,
				``Fractal growth viscous fingers: quantitative characterization of a fluid instability phenomenon'',
				Nature {\bf 314}, 141-144 (1985).

\bibitem{Vicsek}		T. Vicsek, 
				{\it Fractal Growth Phenomena}, 2nd ed. 
				(Singapore, World Scientific, 1992).

\bibitem{Meakin}		P. Meakin,
				{\it Fractals, Scaling and Growth Far from Equilibrium}
				(Cambridge University Press, Cambridge, 1998).

\bibitem{Meakin95}		P. Meakin,
				``Progress in DLA research'',
				Physica D {\bf 86}, 104-112 (1995).

\bibitem{Halsey00}		T. C. Halsey, 
				``Diffusion-limited aggregation: A model for pattern formation'', 
				Phys. Today {\bf 53}, 36-41 (2000).


\bibitem{Sander00}		L. M. Sander,
				``Diffusion-limited aggregation: a kinetic critical phenomenon?'',
				Contemp. Phys. {\bf 41}, 203-218 (2000).


\bibitem{Kesten87}		H. Kesten, 
				``How long are the arms in DLA?'',
				J. Phys. A {\bf 20}, L29 (1987).

\bibitem{Kesten90}		H. Kesten,
				``Upper bounds for the growth rate of DLA'',
				Physica A {\bf 168}, 529--535 (1990).



\bibitem{Muthukumar83}		M. Muthukumar,
				``Mean-Field Theory for Diffusion-Limited Cluster Formation'',
				Phys. Rev. Lett. {\bf 50}, 839 (1983).

\bibitem{Tokuyama84}		M. Tokuyama and K. Kawasaki,
				``Fractal dimensions for diffusion-limited aggregation'',
				Phys. Lett. A {\bf 100}, 337-340 (1984).

\bibitem{Turkevich85}		L. A. Turkevich and H. Scher,
				``Occupancy-Probability Scaling in Diffusion-Limited Aggregation'',
				Phys. Rev. Lett. {\bf 55}, 1026 (1985).


\bibitem{Ball85}		R. C. Ball, R. M. Brady, G. Rossi, and B. R. Thompson,
				``Anisotropy and Cluster Growth by Diffusion-Limited Aggregation'',
				Phys. Rev. Lett. {\bf 55}, 1406 (1985).

\bibitem{Ball86}		R. C. Ball,
				``Diffusion limited aggregation and its response to anisotropy'',
				Physica A {\bf 140}, 62-69 (1986).

\bibitem{Family87}		F. Family and H. G. E. Hentschel,
				``Asymptotic structure of diffusion-limited aggregation clusters in two dimensions'',
				Faraday Discuss. Chem. Soc. {\bf 83}, 139-144 (1987).

\bibitem{Halsey94}		T. C. Halsey,
				``Diffusion-limited aggregation as branched growth'',
				Phys. Rev. Lett. {\bf 72}, 1228 (1994).

\bibitem{Halsey97}		T. C. Halsey, B. Duplantier, and K. Honda,
				``Multifractal Dimensions and Their Fluctuations in Diffusion-Limited Aggregation'',
				Phys. Rev. Lett. {\bf 78}, 1719 (1997).




\bibitem{Meakin85b}		P. Meakin and T. Vicsek,
				``Internal structure of diffusion-limited aggregates'',
				Phys. Rev. A {\bf 32}, 685(R) (1985).

\bibitem{Halsey85}		T. C. Halsey and P. Meakin,
				``Axial inhomogeneity in diffusion-limited aggregation'',
				Phys. Rev. A {\bf 32}, 2546(R) (1985).

\bibitem{Meakin87}		P. Meakin, R. C. Ball, P. Ramanlal, and L. M. Sander,
				``Structure of large two-dimensional square-lattice diffusion-limited aggregates: Approach to asymptotic behavior'',
				Phys. Rev. A {\bf 35}, 5233 (1987).






\bibitem{Meakin85}		P. Meakin, 
				``The structure of two-dimensional Witten-Sander aggregates'', 
				J. Phys. A {\bf 18}, L661 (1985).


\bibitem{Ball85b}		R. C. Ball and R. M. Brady,
				``Large scale lattice effect in diffusion-limited aggregation'',
				J. Phys. A {\bf 18}, L809 (1985).

\bibitem{Meakin86}		P. Meakin,
				``Universality, nonuniversality, and the effects of anisotropy on diffusion-limited aggregation'',
				Phys. Rev. A {\bf 33}, 3371 (1986).

\bibitem{Kertesz86}		J. Kertesz and T. Vicsek,
				``Diffusion-limited aggregation and regular patterns: fluctuations versus anisotropy'',
				J. Phys. A {\bf 19}, L257-L262 (1986).





\bibitem{Halsey86}		T. C. Halsey, P. Meakin, and I. Procaccia,
				``Scaling structure of the surface layer of diffusion-limited aggregates'',
				Phys. Rev. Lett. {\bf 56}, 854 (1986).


\bibitem{Mandelbrot90}		B. B. Mandelbrot and C. J. G. Evertsz,
				``The potential distribution around growing fractal clusters'',
				Nature {\bf 348}, 143 (1990).


\bibitem{Plischke84}		M. Plischke and Z. Racz,
				``Active Zone of Growing Clusters: Diffusion-Limited Aggregation and the Eden Model'',
				Phys. Rev. Lett. {\bf 53}, 415 (1984); 
				see Comment by P. Meakin and L. M. Sander,
				Phys. Rev. Lett. {\bf 54}, 2053 (1985).






\bibitem{Duplantier99}		B. Duplantier,
				``Harmonic Measure Exponents for Two-Dimensional Percolation'',
				Phys. Rev. Lett. {\bf 82}, 3940 (1999).

\bibitem{Smirnov01}		S. Smirnov,
				``Critical percolation in the plane: conformal invariance, Cardy's formula, scaling limits,''
				C. R. Acad. Sci. Paris Sér. I Math. {\bf 333}, 239-244 (2001).

\bibitem{Chelkak12}		D. Chelkak and S. Smirnov,
				``Universality in the 2D Ising model and conformal invariance of fermionic observables'',
				Invent. Math. {\bf 189}, 515-580 (2012).



\bibitem{Stanley77}		H. E. Stanley,
				``Cluster shapes at the percolation threshold: an effective cluster dimensionality and its connection with critical-point exponents'',
				J. Phys. A: Math. Gen. {\bf 10}, L211 (1977).

\bibitem{Mandelbrot}		B. B. Mandelbrot, 
				{\it The Fractal Geometry of Nature}
				(San Francisco, Freeman, 1982).




\bibitem{Loh14}			Y. E. Loh, 
				``Bias-free simulation of diffusion-limited aggregation on a square lattice'',
				ArXiv 1407.2586v1 (2014).


\bibitem{Mandelbrot02}		B. B. Mandelbrot, B. Kol, and A. Aharony,
				``Angular Gaps in Radial Diffusion-Limited Aggregation: Two Fractal Dimensions and Nontransient 
				Deviations from Linear Self-Similarity'',
				Phys. Rev. Lett. {\bf 88}, 055501 (2002).

\bibitem{Tolman89}		S. Tolman and P. Meakin,
				``Off-lattice and hypercubic-lattice models for diffusion-limited aggregation in dimensionalities 2-8'',
				Phys. Rev. A {\bf 40}, 428 (1989).





\bibitem{Vicsek90}		T. Vicsek, F. Family, and P. Meakin,
				``Multifractal geometry of diffusion-limited aggregates'',
				Eur. Phys. Lett. {\bf 12}, 217 (1990).

\bibitem{Ossadnik91}		P. Ossadnik, 
				``Multiscaling Analysis of Large-Scale Off-Lattice DLA'', 
				Physica A {\bf 176}, 454 (1991).


\bibitem{Adams09}		D. A. Adams, L. M. Sander, E. Somfai, and R. M. Ziff,
				``The harmonic measure of diffusion-limited aggregates including rare events'',
				Eur. Phys. Lett. {\bf 87}, 20001 (2009).

\bibitem{Hanan12}		W. G. Hanan and D. M. Heffernan,
				``Multifractal analysis of the branch structure of diffusion-limited aggregates'',
				Phys. Rev. E {\bf 85}, 021407 (2012).


\bibitem{Menshutin12}		A. Menshutin,
				``Scaling in the Diffusion Limited Aggregation Model'',
				Phys. Rev. Lett. {\bf 108}, 015501 (2012).



\bibitem{Komlos76}		J. Koml\'os, P. Major, and G. Tusn\'ady,
				``An approximation of partial sums of independent RV's, and the sample DF. II.''
				Z. Wahrscheinlichkeitstheorie verw. Gebiete {\bf 34}, 33-58 (1976).

\bibitem{Kozdron05}		M. Kozdron and G. Lawler,
				``Estimates of random walk exit probabilities and application to loop-erased random walk'',
				Electr. J. Probab. {\bf 10}, 1442-1467 (2005).



\bibitem{Bogoyavlenskiy02}	V. A. Bogoyavlenskiy, 
				``How to grow isotropic on-lattice diffusion-limited aggregates'',
				J. Phys. A: Math. Gen. {\bf 35}, 2533 (2002).

\end{thebibliography}
\end{document}